\documentclass[twocolumn,showpacs,preprintnumbers,amsmah,amssymb]{revtex4}
\usepackage{graphicx}% Include figure files
\usepackage{dcolumn}% Align table columns on decimal point
\usepackage{bm}% bold math
 \usepackage[latin1]{inputenc}

\begin{document} % INITIALIZE - DONT CHANGE

\title{Training Induced Positive Exchange Bias  in NiFe/IrMn Bilayers}

\author{ S. K. Mishra, F. Radu$^*$, H. A. D\"urr and W. Eberhardt}
 \affiliation{ Helmholtz-Zentrum Berlin f\"ur Materialien und
Energie, Albert-Einstein-Str. 15, D-12489 Berlin, Germany}
\date{\today}
\begin{abstract}
Positive exchange bias  has been observed in the
Ni$_{81}$Fe$_{19}$/Ir$_{20}$Mn$_{80}$ bilayer system via soft x-ray
resonant magnetic scattering. After field cooling of the system through the blocking
temperature of the antiferromagnet, an initial conventional negative
exchange bias is removed after training i. e.
successive magnetization reversals, resulting in a positive exchange bias
for a temperature range down to 30 K below the blocking temperature (450~K).
This new manifestation of magnetic training is discussed in terms of metastable magnetic disorder
at the magnetically frustrated interface during  magnetization reversal.

\end{abstract}
 \maketitle

The exchange bias in a ferromagnetic (FM) /antiferromagnetic (AF)
system was first discovered by Meiklejohn and Bean
~\cite{bean:1957} in Co particles encapsulated by a shell of
antiferromagnetic CoO.  For more than 60 years this effect has
been extensively studied, mainly due to the elusiveness of a
fundamental understanding and its value for applications such as
ultrahigh-density magnetic recording, giant magnetoresistance
(GMR), and spin valve devices~\cite{dieny:1999}. When a sample
with a magnetically uncompensated  FM/AF interfaces is cooled
through the N\'eel temperature (T$_{N}$) of the AF, with the Curie
temperature (T$_{C}$) of the FM being higher than T$_{N}$, an
unidirectional exchange anisotropy, namely, exchange bias (EB) is
induced in the system~\cite{berkowitz:1999,nogues:1999,radu:2008}.

Usually the exchange bias direction is opposite (negative EB)  to
the FM magnetization direction during field cooling. The reverse
situation, namely a shift of the hysteresis loop to positive
direction (positive EB) with respect to field cooling directions
occurs too.  Positive exchange bias (PEB)  was first observed in
FeF$_{2}$ /Fe bilayers and was associated with antiferromagnetic
interfacial coupling~\cite{nogues:1996} which was  recently
observed experimentally~\cite{roy:2005,ohldag:2006}. In these
systems the magnitude and sign of EB depends strongly on strength
and direction of the cooling field (H$_{CF}$)~\cite{nogues:1996}.
This is in agreement with the results obtained by Leighton
\emph{et. al.} for MnF$_{2}$/Fe bilayers~\cite{leighton:2000}.
Beckmann and Usadel~\cite{beckmann:2006} found
%studied the relation between EB and the direction of H$_{CF}$
using Monte Carlo simulations that a directional variation of
H$_{CF}$ can even result in different  asymmetric magnetization
reversal modes of the FM. It was also reported that a diluted
AF order at the interface may enhance the EB~\cite{miltenyi:2000},
and that the spin alignment at FM/AF interfaces~\cite{tsai:2003}
depends on their roughness.

Another category of  PEB has been observed for instance in
Cu$_{1-x}$Mn$_{x}$/Co and CoO/Co bilayers, where PEB is
established only in the proximity of the blocking temperature,
T$_{B}$, ~\cite{gredig:2002,radu:2003,kohlhepp:2007,ali:2007}
after field cooling. Further lowering of the sample temperature
results in negative exchange bias. The microscopic mechanism of
the PEB close to T$_{B}$ is discussed on the basis of coexistence
of FM and AF interface coupling~\cite{radu:2003}, interfacial RKKY
coupling for the Cu$_{1-x}$Mn$_{x}$/Co bilayer~\cite{ali:2007},
and unidirectional coercivity
enhancement~\cite{gredig:2002,kohlhepp:2007}. In general, the
phenomena of PEB close to T$_{B}$ is only observed for finite
magnetic bilayer thicknesses while the microscopic origin, the
impact of H$_{CF}$ and competing coupling mechanisms at the
interface are still debated.

Here we report on a new  manifestation of exchange bias
in IrMn based EB bilayers,  which is one of the most attractive AF
material for both device applications and fundamental
research~\cite{camarero:2005, radu:jpcm:2006,
ohldag:2003,tsunoda:2006,steenbeck:2007, suess:2003,mccord:2003}.
 We show that PEB in the Ni$_{81}$Fe$_{19}$/Ir$_{20}$Mn$_{80}$ bilayers is
  induced in our samples only after several training cycles near T$_{B}$. This is
different from all previous cases where PEB is already observed
for the very first hysteresis loop after field cooling. Moreover,
a new type of asymmetric magnetization reversal can be inferred by
analyzing the shape of the hysteresis loop. The experimental
results are discussed in the framework of frustration at the
magnetically disordered AF/FM interface.

%\newpage
A polycrystalline Si (100)/Cu (5 nm)/ Ni$_{81}$Fe$_{19}$(7.5 nm)/
Ir$_{20}$Mn$_{80}$ (3.5 nm)/ Cu (2.5 nm) sample was grown by
magnetron sputtering on a Si wafer at a base pressure of 8.3
$\times$ 10$^{-9}$ mbar. Ultra clean Ar gas was used as the
sputtering medium. The partial Ar pressure during growth was 1.5
$\times$ 10$^{-3}$ mbar. An uniaxial magnetic anisotropy was
induced in the FM layer by applying an in-situ magnetic field of 2
KOe parallel to the sample surface. A Cu~(5~nm) buffer layer was
deposited  to promote a smooth growth of the magnetic
hetrostructure. The subsequent Ni$_{81}$Fe$_{19}$ (Py) and
Ir$_{20}$Mn$_{80}$ layers were capped with Cu~(2.5~nm) to prevent
oxidation of the hetrostructure.

X-ray resonant magnetic scattering measurements were performed at
BESSY II in the ALICE diffractometer~\cite{grabis:2003} installed
at the BESSY II PM3 bending magnet  beamline. Magnetic hysteresis
loops for the ferromagnetic layer were measured by tuning the
x-ray energy into the Ni L$_3$ absorption edge and monitoring the
specularly reflected x-ray intensity as a function of magnetic
field applied parallel to  the sample surface and in the
scattering plane. Maximum magnetic sensitivity  i.e. asymmetry,
(I$^{+}$ - I$^{-}$)/(I$^{+}$ + I$^{-}$),( where; I$^{\pm}$ are
reflected intensity for opposite magnetic field directions) was
achieved by utilizing 80\% circularly polarized x-rays in specular
condition at an incident angle equal to  $\theta$ = 9.9°.

\begin{figure}[ht]
\includegraphics[clip=true,keepaspectratio=true,width=.8\linewidth]{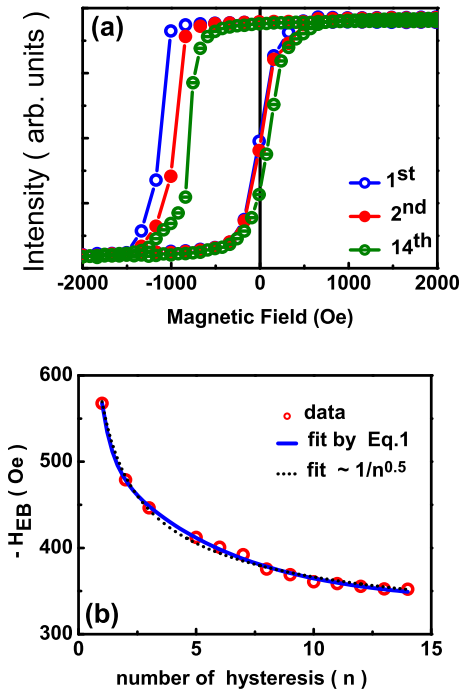}
      \caption {(color online)(a) Magnetization curves  measured at 10~K
         after field cooling ($H_{FC}$=2~KOe) the system from 470~K through the
         blocking temperature. The 1$^{st}$ (blue), 2$^{nd}$ (red) and
         14$^{th}$(green) hysteresis loops are shown.(b) H$_{EB}$ as a function of the loop
          index (n) extracted from individual hysteresis loops. Open circles
         are the experimental data, the  line represents a model, and the dotted line is a $1/\sqrt{n}$ functional fit (see text).}
          %, (c) H$_{C}$ of NiFe/IrMn exchange biased polycrystalline bilayer as the function
          %of the number index (n) of the hysteresis cycles. Points corresponds to the experimental data
          %whereas the  line represent to the fit.
        \label{fig1}
         \end{figure}

Fig.~\ref{fig1}a shows the magnetic training effect observed in
the Ni$_{81}$Fe$_{19}$(7.5 nm)/Ir$_{20}$Mn$_{80}$ (3.5 nm) sample
at 10 K after field cooling from T=470~K through the blocking
temperature, T$_{B}$=450~K. The first hysteresis loop exhibits a
sharp reversal at H$_{c1}$ (the coercive field at the very first
reversal) and a more rounded reversal at H$_{c2}$ (the coercive
field at the second reversal).(see fig. 1a) By measuring a second
hysteresis loop we observe a decrease of the exchange bias field,
H$_{EB}$= (H$_{c1}$ - H$_{c2}$)/2, which is characteristic for a
training effect. Strikingly, the second hysteresis loop exhibits
the same steep/rounded characteristic features at
H$_{c1}$/H$_{c2}$ as the first one. Even after 14th hysteresis  a
sharper reversal at H$_{c1}$ as compared to H$_{c2}$ is preserved.
This training effect  is different as the one observed for the
archetypal Co/CoO EB bilayer~\cite{gredig:2002, radu:2002,
gruyters:2000, brems:2005} and also for IrMn/CoFe bilayers
\cite{mccord:2003}. There, at the very first reversal after field
cooling a transition from an essentially single AF domain to a AF
multidomain state occurs which leads to a transition from a
pronounced asymmetric hysteresis to an essentially symmetric
one~\cite{radu:2002}. The shape of the hysteresis at the second
reversal was more rounded, and in contrast to our current
observation, the consecutive hysteresis loops remain essentially
rounded at both hysteresis loop branches. Such a transition to a
symmetric hysteresis behavior is characteristic for changes in the
bulk AF domain structure. Since we do not observe this behavior in
our system we believe  that  the AF bulk spin structure  is
robust, lacking a dramatic change of the spin configurations which
would naturally  lead to a pronounced change of loop asymmetry.
Also, higher orders anisotropy of the AF
layer~\cite{hoffmann:2004,lund:2007} may not be the main origin of
this type of training effect, since the AF layer exhibits an
uniaxial anisotropy, as confirmed by azimuthal dependence of the
coercive fields (data not shown).

In order to explain the asymmetry of the magnetization curve in
Fig.~\ref{fig1}a we assume that the AF layer behaves virtually as
predicted by the Meiklejohn and Bean (M\&B) model. Only when the
AF thickness is slightly larger than a critical value, the AF
spins rotate reversible away from their stable angular
orientations set by a field cooling procedure. During the
magnetization reversal,  the AF spin direction acquires a maximum
value of 45° at the critical thickness~\cite{radu:2008}. When this
angular deviation is significant, an asymmetric magnetization
reversal occurs~\cite{radu:2008}. Note that the condition of being
close to the critical thickness regime is realized in our system.
The  critical AF thickness for exchange bias was measured (data
not shown) and is about 2~nm. Using a modified M\&B model named
Spin Glass (SG) model~\cite{radu:2008} this asymmetry of the
hysteresis loop is reproduced numerically at reduced thicknesses.
An enhanced coercivity is also accounted for by assuming a
magnetically disordered interface (see Fig. 3.39 in
\cite{radu:2008}).

The  stability of the bulk AF structure is seen also when plotting
the exchange bias field, H$_{EB}$,  as a function of hysteresis
loop index $n$ in Fig.~\ref{fig1}b.  We find that the EB field as
a function of n can be described by the following empirical
law~\cite{paccard:1966}: $|H_{EB}|=|H_{EB}^\infty|+k/\sqrt{n}$,
with $H_{EB}^\infty=272\pm4$~Oe and $k=297\pm7$~Oe. A less
significant decrease of the EB takes place between the first and
the second hysteresis cycles, suggesting that no AF domains
(rearrangements) occur. A monotonous  evolution of the coercivity
and H$_{EB}$ as a function of n appears due to the interfacial
spin rearrangement at the magnetically disordered FM/AF interface.
The presence of interfacial spin frustration can enhance the
interface area remarkably while keeping the total spin number
preserved. At the FM/AF interface the AF magnetic anisotropy is
assumed to be modified, leading to essentially two different types
of AF uncompensated spins after field cooling: namely frozen and
rotatable AF uncompensated spins being rigidly exchange coupled to
the AF and FM layers, respectively~\cite{radu:2008}.

With each cycle a spin rearrangement takes place and this modifies
the coercive and exchange bias fields. Note that our approach is
different as the one of Binek ~\cite{binek:2004}, although the
main concept of interfacial magnetic instabilities is preserved.
While Binek considers only a change of the interfacial AF
magnetization, we suggest that both components, frozen and
rotating are affected by the FM magnetization reversals. Moreover,
mixed ferromagnetic and antiferromagnetic coupled components will
contribute distinctively,  through different relaxations rates, to
the training effect.

Additional evidence for this scenario can be obtained by describing the
data in Fig.~\ref{fig1}b with the following expression
to simulate the relaxation of the exchange bias as a function of $n$:
\begin{equation}
H _{EB}^{n}  = H_{EB}^{\infty} + A_{f} exp(-n/P_{f}) + A_{i}
exp(-n/P_{i}), \end{equation} We note that it is not possible to
describe the curve in Fig.~\ref{fig1}b by only one exponential.
Here, $H _{eb}^{n}$ is the exchange bias of the $n^{th}$
hysteresis loop, A$_{f}$ and P$_{f}$ are parameters related to the
change of the frozen spins, A$_{i}$ and P$_{i}$ are evolving
parameters of the interfacial magnetic frustration of the bilayer.
The A parameters have dimension of magnetic field (Oersted) while
the P's are dimensionless parameters and resemble a relaxation
time, where the continuous variable is replaced by a discrete
variable, namely the hysteresis index $n$. The parameters obtained
from  fit to the H$_{EB}$ data are: $H_{EB}^{\infty}$ =
335${\pm}$5  Oe, A$_{f}$ = 641${\pm}$527  Oe,
 P$_{f}$ = 0.44${\pm}$ 0.18, A$_{i}$ = 199${\pm}$13  Oe, P$_{i}$ = 5${\pm}$0.6.

Indeed, within the SG approach, we distinguish a sharp
contribution due to  uncompensated spins at the interface and a
much weaker decrease from the frozen uncompensated spins. The
frozen component appears to relax about 10 times slower as
compared to the other one.

\begin{figure}[ht]
\includegraphics[clip=true,keepaspectratio=true,width=.8\linewidth]{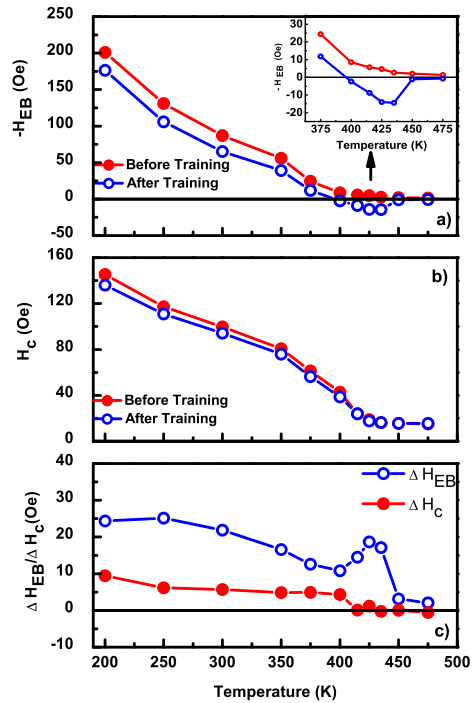}
         \caption{(color online)Temperature dependence of the (a)
         H$_{EB}$ for the first (red filled circles) and the last loop (blue open circle),
         (b) H$_{c}$  for the first (red filled circles) and the last loop (blue open circle),
         and (c) difference of magnitude (before and
         after training) for H$_{EB}$ and H$_{c}$ fields
         of the FM layer. First loop is measured right after
         field cooling whereas the last loop was measured after  fast
         flipping of magnetic field. The inset in Fig. (a) is an
         enlargement at the blocking temperature showing the positive exchange bias.}
         \label{fig2}
         \end{figure}

\begin{figure}[ht]
\includegraphics[clip=true,keepaspectratio=true,width=1\linewidth]{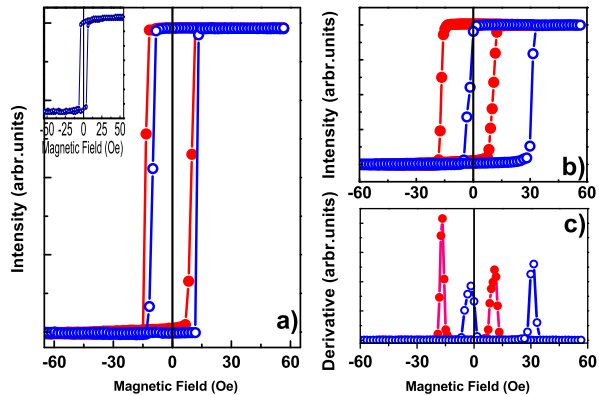}
          \caption{(color online)The MH loops of the FM for  two representative
          temperatures T=450~K (panel a)  and T=435~K (panel b), respectively,
          measured before (red filled circle) and after (blue open circle)
          training. The system has been field cooled ($H_{FC}$=2~KOe) from T=470~K.
          The inset shows  the hysteresis of a Ni$_{81}$Fe$_{19}$(7.5 nm) layer
          in the absence of a IrMn layer. In panel c) the
          derivative of the hysteresis loops (of panel b)).
          The asymmetry of each hysteresis and its reversal are clearly seen as a
          different amplitudes and widths at the coercive fields when comparing the
          ascending and the descending branches. }
          \label{fig3}
 \end{figure}

For the remainder of this letter we demonstrate the importance of
magnetic training for establishing PEB. The temperature dependence
of the H$_{EB}$ and H$_{c}$ for the bilayer is shown schematically
in Fig.~\ref{fig2}. The data were obtained by field cooling the
sample in an external magnetic field of about 2~KOe from T=470~K
through the blocking temperature at each temperature shown in
Fig.~\ref{fig2}. Even up to the highest available temperatures in
our experimental setup~(470~K) the measured coercivity values were
higher than that of a single Py layer (H$_c$(Py)$\approx$~5~Oe)
(see Fig.~\ref{fig3}a). This indicates that the N\'{e}el
temperature of the AF film was not reached. The data displayed in
Fig.~\ref{fig2} were obtained from hysteresis loops of the freshly
biased system (red lines and open symbols) and after training via
30 hysteresis sweeps (blue lines and solid symbols). The
hysteresis loops in Fig.~\ref{fig2}a display a roughly linear
increase of H$_{EB}$ with decreasing temperature.

Interestingly, the exchange bias after training is almost rigidly
shifted to lower values over the whole measured temperature
region. This can be clearly seen in Fig.~\ref{fig2}c where the
difference in exchange bias, $\Delta
H_{EB}=H_{EB}(n=1)-H_{EB}(n=30)$, before and after training is
shown (blue lines and solid symbols). This decrease of the EB
after training appears to to contribute to  the PEB near T$_{B}$
(see the inset of Fig.~\ref{fig2}a). However, the size of the PEB
is actually larger than that expected from the rigid shift at
lower temperatures. Another  new observation here is that the PEB
occurs only after training. For instance,  at T=435~K a typical
negative exchange bias extracted from the very first hysteresis
cycle after cooling has changed sign, thus, resulting in PEB after
training.

We can discriminate between several possible mechanisms leading to
PEB. We observe a rather constant temperature dependence for the
$\Delta H_{EB}$ in Fig.~\ref{fig2}c. A dominant RKKY
origin~\cite{ali:2007} of the PEB would lead to a nonmonotonic
temperature dependence for the $\Delta H_{EB}$, which is not
observed. Also, an unidirectional enhancement of coercivity as a
main reason for  an apparent PEB~\cite{gredig:2002,kohlhepp:2007}
is not fully supported by our data. A rotation of bulk AF grains
or domains would have to be suppressed at low temperatures. Our
data shows no significant exchange bias and coercivity variation
($\Delta H_{EB}$ and $\Delta H_{c}$) across the T$_B$ and closely
below it. As a result we are lead to the conclusion that
irreversible changes at the interface are responsible for the PEB.

In order to explain the occurrence of PEB we assume a simple model
based on the previous description of PEB~\cite{nogues:1996,
radu:2003}, where an uncompensated spin component exhibiting a
fundamentally antiparallel coupling to the FM is needed.

During the field cooling procedure a minority of the uncompensated
interfacial AF spins will prefer to align antiparallel to the
direction of the FM layer defined by the cooling field. This
situation would result in a typical negative
EB~\cite{nogues:1996}. After training, this minority component
will rotate irreversibly due to consecutive magnetization
reversals of the FM acting on a frustrated spin state. This
frustrated spin state is a consequence of symmetry breaking at the
FM/AF interfaces~\cite{kuch:2006,krug:2008}. As a result, a weak
positive shift of the hysteresis loop will occur at all
temperatures after training. When, cooling from above T$_B$, this
component will dominate strongly right below the blocking
temperature since it is anchored stronger to the bulk side of the
AF layer. Only for lower temperatures, the majority component
providing negative exchange bias will lead to a robust exchange
bias.

The rotation of this minority component is clearly seen in
Fig.~\ref{fig3}b) and Fig.~\ref{fig3}c). The untrained hysteresis
loop at T=435~K is asymmetric, namely the first reversal is
steeper as compared to the second one (compare the density of
field points across the reversals at the coercive fields). After
training, this asymmetry reverses, namely the second reversal
becomes steeper. This is a direct proof of that a minority
unidirectional anisotropy responsible for  the PEB has rotated
during field cycling.

In conclusion,  we have observed a novel asymmetry of the
hysteresis loop predicted numerically at the critical region for
exchange bias. Training effect leads to  irreversible changes of
an essentially frustrated interfacial spin state. At the blocking
temperature a positive exchange bias occurs after training effect.
A rotation of a minority antiparallel coupled  spin component is
clearly revealed through the asymmetric nature of the hysteresis
loops. The experimental data allows to discriminate between
different models for the newly observed positive exchange bias,
supporting a mixture of antiferromagnetic (minority) a
ferromagnetic (majority) coupling mechanism at the interface.

We gratefully acknowledge Dr. T. Kachel  and Dr. R. Follath for
providing excellent  support during the measurement at PM3
(BESSY). The ALICE diffractometer  is funded through the BMBF
contract No. 05KS7PC1.

%\newpage


\begin{thebibliography}{99}


\bibitem{bean:1957}
W. Meiklejohn, and C. P. Bean, Phys. Rev. \textbf{105}, 904 (1957).

\bibitem{dieny:1999} B. Dieny, J. Magn. Magn. Mater. \textbf{136},
335 (1994).

\bibitem{berkowitz:1999}


Berkowitz et al.,
%A. E. Berkowitz, and K. Takano,
J. Magn. Magn. Mater., \textbf{200},552-570 (1999).


\bibitem{nogues:1999}

Nogu\'es et al.,
%J. Nogu\'es, and I. K. Schuller,
J. Magn. Magn. Mater. \textbf{192}, 203 (1999).

\bibitem{radu:2008}
F. Radu  and H. Zabel, Springer Tracts in Modern Physics,
\textbf{227}, 97 (2008).

\bibitem{nogues:1996}
 Nogu\'es et al.,
%J. Nogu\'es, D. Lederman, T. J. Moran, and I. K. Schuller,
 Phys. Rev. Lett. \textbf {76}, 4624 (1996).

\bibitem{roy:2005}
Roy et al., Phys. Rev. Lett. \textbf{95}, 047201 (2005).

\bibitem{ohldag:2006}
Ohldag et al.,
 Phys. Rev. Lett. \textbf{96}, 027203 (2006).

%\bibitem{tang:2006}
%Tang et al.,
%Y. K. Tang, Y. Sun, and Zh. H. Cheng,
%J. Appl. Phys. \textbf {100},023914 (2006).

%\bibitem{yang:2005}
%Yang et al.,
% D. Z. Yang, J. Du, L. Sun, X. S. Wu, X. X. Zhang, and S. M. Zhou,
%Phys. Rev. B \textbf {71}, 144417 (2005).

\bibitem{leighton:2000}
Leighton et al.,
%C. Leighton, J. Nogu\'es, B. J. J\"onsson-\AA kerman, and I. K.
%Schuller,
Phys. Rev. Lett. \textbf {84}, 3466 (2000).

\bibitem{beckmann:2006}
Beckmann et al.,
%B. Beckmann and K. D. Usadel,
 Phys. Rev. B \textbf {74}, 054431(2006).


\bibitem{miltenyi:2000}
Milt\'enyi et al.,
% P. Milt\'enyi, M. Gierlings, J. Keller, B. Beschoten, G. G\"untherodt,
  %U. Nowak, and K. D. Usadel,
  Phys. Rev. Lett. \textbf {84}, 4224 (2000).

\bibitem{tsai:2003}
Tsai et al.,
% S. H. Tsai, D. P. Landau, and T. C. Schulthess,
 J. Appl. Phys. \textbf {93}, 8612 (2003).


\bibitem{gredig:2002}
Gredig et al.,
%T. Gredig, I. N. Krivorotov, P. Eames and D. Dahlberg,
Appl. Phys. Lett. \textbf{81}, 1270 (2002).

\bibitem{radu:2003}
Radu et al.,
 %F. Radu, M. Etzkorn, R. Siebrecht, T. Schmitte, K. Westerholt, and H. Zabel,
 Phys. Rev. B \textbf {67}, 134409 (2003).



\bibitem{kohlhepp:2007}
Kohlhepp et al.,
%J. T. Kohlhepp,  H. Wieldraaijer and  W. J. M. de Jonge,
J. Mat. Res. \textbf{22}, 569 (2007).


\bibitem{ali:2007}
Ali et al.,
%M. Ali, P. Adie, C. H. Marrows, D. Greig, B. J. Hickey, and R. L.
%Stamps,
Nature Materials, \textbf {6}, 70 (2007).


\bibitem{camarero:2005}
Camarero et al.,
%J. Camarero, J. Sort and A. Hoffmann, J. Miguel
%Garc{\'i}a-Mart{\'i}n, B. Dieny, R. Miranda and Josep Nogu{\'e}s,
Phys. Rev. Lett. \textbf{95}, 057204 (2005).



\bibitem{radu:jpcm:2006}
Radu et al.,
%F. Radu, A. Westphalen, K. Theis-Br\"ohl, and H. Zabel.
J. Phys.:Condens. Matter \textbf{18}, L29-L36, (2006).

\bibitem{ohldag:2003}
Ohldag et al.,
% H. Ohldag, A. Scholl, F. Nolting, E. Arenholz,
%S. Maat, A. T. Young, M.Carey, and J. St\"ohr,
Phys. Rev. Lett. \textbf{91}, 017203 (2003).

\bibitem{tsunoda:2006}
Tsunoda et al.
% T. Nakamura, M. Naka and S. Yoshitaki, C. Mitsumata, and M. Takahashi
Appl. Phys. Lett. \textbf{89}, 172501(2006).

\bibitem{steenbeck:2007}
Steenbeck et al.,
%K. Steenbeck, and R. Mattheis,
J. Magn. Magn. Mater. \textbf{316},90 (2007).


\bibitem{suess:2003}
Suess et al.,
   %D. Suess, M. Kirschner, T. Schrefl, J. Fidler, R. L. Stamps and J.-V. Kim,
   Phys. Rev. B  \textbf{67}, 054419 (2003).

\bibitem{mccord:2003}
McCord et al.,
 %%J. McCord,  R. Mattheis and R. Schäfer,
 J. Appl. Phys.\textbf{93},5491 (2003).

\bibitem{grabis:2003}
  Grabis et al.,
  %J. Grabis, A. Nefedov and H. Zabel,
   Rev. Sci. Instr.  \textbf{74}, 4048 (2003).

\bibitem{radu:2002}
Radu et al.,
%F. Radu, M. Etzkorn, T. Schmitte, R. Siebrecht, A. Schreyer, K.
%Westerholt, and H. Zabel,
J. Magn. Magn. Mater. \textbf{240}, 251 (2002).


\bibitem{gruyters:2000}
M. Gruyters and D. Riegel,  Phys. Rev. B  \textbf{63}, 052401 (2000).

\bibitem{brems:2005}
Brems et al.,
%S. Brems, D. Buntinx, K. Temst, and C. Van
%Haesendonck, F. Radu and H. Zabel,
Phys. Rev. Lett. \textbf{95}, 157202 (2005).

\bibitem{hoffmann:2004}
A, Hoffmann, Phys. Rev. Lett. \textbf{93}, 097203 (2004).


\bibitem{lund:2007}
M. S. Lund, C. Leighton, Phys. Rev. B  \textbf{76}, 104433 (2007).

\bibitem{binek:2004}
   Ch. Binek, Phys. Rev. B  \textbf{70}, 014421 (2004).

%\bibitem{ali:2003}
%Ali et al.,
 %M. Ali, C. H. Marrows, M. Al-Jawad,  B. J. Hickey, A. Misra, U. Nowak,
 %and K. D. Usadel,
 %Phys. Rev. B \textbf{68}, 214420 (2003).
\bibitem{paccard:1966}
D. Paccard, et al., Phys. Status Solidi \textbf{16}, 301 (1966).

\bibitem{kuch:2006}
Kuch et al., Nature Materials,  \textbf{5}, 128 (2006).


\bibitem{krug:2008}
Krug et al., %I. P. Krug,  F. U. Hillebrecht, M. W. Heverkort, A.
%Tanaka, L. H. Tjeng, H. Gomonay, A. Fraile-Rodriguez, F. Nolting,
%S. Cramm, and C. M. Scneider,
Phys. Rev. B \textbf{78}, 064427 (2008).


\end{thebibliography}
\end{document}